\documentclass[conference]{IEEEtran}
\IEEEoverridecommandlockouts

\usepackage{cite}
\usepackage{amsmath,amssymb,amsfonts}
\usepackage{algorithmic}
\usepackage{graphicx}
\usepackage{textcomp}
\usepackage{xcolor}

\usepackage{hyperref}
\hypersetup{
    colorlinks=true,
    citecolor=blue,
}
\usepackage{colortbl}

\usepackage{cite}
\usepackage{url}
\usepackage{amsfonts}
\usepackage{multirow}
\usepackage{makecell}
\usepackage{booktabs} 
\usepackage{subcaption}

\usepackage{pifont}

\begin{document}

\title{The AudioMOS Challenge 2025
\thanks{EC was supported by JSPS KAKENHI Grant Number JP25K00143.}
}

\author{
    \IEEEauthorblockN{
        \IEEEauthorblockA{\textit{
            Wen-Chin Huang\IEEEauthorrefmark{1}, %
            Hui Wang\IEEEauthorrefmark{2}, %
            Cheng Liu\IEEEauthorrefmark{2}, %
            Yi-Chiao Wu\IEEEauthorrefmark{3}, %
            Andros Tjandra\IEEEauthorrefmark{3},}}
        \IEEEauthorblockA{\textit{
            Wei-Ning Hsu\IEEEauthorrefmark{3}, %
            Erica Cooper\IEEEauthorrefmark{4}, %
            Yong Qin\IEEEauthorrefmark{2}, %
            Tomoki Toda\IEEEauthorrefmark{1}}}
    }
    \IEEEauthorblockA{
        \IEEEauthorrefmark{1}Nagoya University, Japan
    }
    \IEEEauthorblockA{
        \IEEEauthorrefmark{2}Nankai University, China
    }
    \IEEEauthorblockA{
        \IEEEauthorrefmark{3}Meta, USA
    }
    \IEEEauthorblockA{
        \IEEEauthorrefmark{4}National Institute of Information and Communications Technology, Japan\\
        Email: voicemos.challenge@gmail.com\\
        \url{https://sites.google.com/view/voicemos-challenge/audiomos-challenge-2025}
    }
}

\maketitle

\begin{abstract}
This is the summary paper for the AudioMOS Challenge 2025, the very first challenge for automatic subjective quality prediction for synthetic audio. The challenge consists of three tracks. The first track aims to assess text-to-music samples in terms of overall quality and textual alignment. The second track is based on the four evaluation dimensions of Meta Audiobox Aesthetics, and the test set consists of text-to-speech, text-to-audio, and text-to-music samples. The third track focuses on synthetic speech quality assessment in different sampling rates. The challenge attracted 24 unique teams from both academia and industry, and improvements over the baselines were confirmed. The outcome of this challenge is expected to facilitate development and progress in the field of automatic evaluation for audio generation systems.
\end{abstract}


\section{Introduction}

With the rapid progress of generative AI technologies, there has been an increasing effort in developing audio generation systems to synthesize speech, singing voices, music, or even sound effects. It should be noted that the evaluation of synthetic audio is a task of equal importance to the development of the generation systems themselves. As the end users of the audio generation systems are often human, it is thus crucial to rely on listening tests like the mean opinion score (MOS) test to conduct such an evaluation. However, subjective tests can be limited by budget constraints, leading to recent interest in the development of automatic assessment methods that are efficient and capture human preferences well at the same time.

An attempt to promote the field of automatic \textit{speech} subjective quality assessment is the VoiceMOS Challenge (VMC) series. Founded in 2022, this annual challenge aimed to use standardized datasets in diverse and challenging domains to understand and compare prediction techniques for human ratings of speech, and furthermore to foster development in this field \cite{voicemos2022,voicemos2023,voicemos2024}. The first VMC was based on the BVCC dataset \cite{bvcc} with an in-domain setting, which is a collection of human-rated speech samples from 187 text-to-speech (TTS) and voice conversion systems. The succeeding versions expanded the scope to a fully zero-shot setting, and tested on cross-lingual TTS, singing voice generation systems, and even noisy speech. The VMC series attracted 40 participating teams over the three-year span, and successfully promoted and progressed the field.

On the other hand, we observe a relatively lagged development in evaluation methods for audio modalities other than speech, especially music and general audio. 
The most commonly adopted evaluation metric is the Fréchet Audio Distance (FAD) \cite{fad}, which compares the statistics between the set of generated samples and a reference set in an embedding space. In the context of text-to-music (TTM) and text-to-audio (TTA) generation, the CLAP Score \cite{make-an-audio} is also commonly adopted, which calculates the cosine similarity between CLAP embeddings of the input text (also known as the \textit{prompt}) and the output. However, studies have shown that such metrics can be sensitive to sample size and the embedding \cite{fad-correlates-poorly, fad-is-embed-sensitive}, while correlating poorly with human perception \cite{fad-correlates-poorly}.
We suspect the reason to be that the evaluation of these two modalities is considered more difficult: in TTM and TTA, the number of valid realizations of a given text input (also known as the \textit{prompt}) can be infinite.
It would therefore be desirable to develop evaluation methods that directly learn from human labels. Such annotated datasets, nonetheless, do not yet exist.

\begin{table*}[t]
	\centering
	\caption{Summary of the datasets for each track.}
	
	\centering
	\begin{tabular}{ c c c c c c c c c}
		\toprule
		\multirow{2}{*}[-1pt]{Track} & \multirow{2}{*}[-1pt]{Type} & \multicolumn{3}{c}{\# Samples} & \multicolumn{3}{c}{\# Systems (conditions)} & \multirow{2}{*}[-2pt]{\makecell{\# ratings\\per sample}}\\
		\cmidrule(lr){3-5} \cmidrule(lr){6-8}
		& & Train & Dev & Test & Train & Dev & Test & \\
		\midrule
		1 & TTM & 1923 & 412 & 413 & 31 & 29 & 30 & 5 \\
		\midrule
		2 & TTS, TTA, TTM & 2700 & 250 & 3060 & \multicolumn{2}{c}{N/A$^\dagger$} & 36 & 10\\
            \midrule
            3 & Synthetic speech in different sampling frequencies &  \multicolumn{2}{c}{400$^\ddagger$} & 400 & \multicolumn{2}{c}{20$^\ddagger$} & 20 & 10 \\
		\bottomrule
            \multicolumn{9}{l}{$^\dagger$: The training set of track 2, the AES-natural dataset, consisted of natural samples, thus the concept of ``system'' and ``condition'' does not apply.} \\
            \multicolumn{9}{l}{
                \begin{tabular}[l]{@{}l@{}}
                    $^\ddagger$: For track 3, the samples (and thus the systems) are shared for training and development, with labels from different listening tests. \\ ~~~~See Sec.~\ref{ssec:track3} for details.
                \end{tabular}
            }
	\end{tabular}
	\label{tab:datasets}
\end{table*}

In light of this, we organized the AudioMOS Challenge (AMC) 2025, an expanded version of the previous VMC series.
The AMC 2025 consists of three tracks.
The first track focused on MOS prediction of TTM systems.
The second track was based on Meta Audiobox Aesthetics \cite{aes}, a suite of unified assessment methods for speech, music, and sound.
The third track focused on MOS prediction for synthesized speech in different sampling frequencies.
The description of each track will be presented in Section~\ref{sec:description}.
In total, from academia and industry, \textbf{24} unique teams submitted their final predictions, and there were 9/10/7 submissions for tracks 1/2/3, respectively. Each track also came with a baseline system, and a brief description will be presented in Section~\ref{sec:participants}. From the results shown in Section~\ref{sec:results}, we are pleased to find that the baseline in each track was outperformed by most participating teams, demonstrating that the challenge indeed fostered advancements. 
In Section~\ref{sec:system-description-forms}, we summarize the system description forms submitted by participants, identifying common and successful techniques by top systems and future directions pointed to directly by the research community.

\section{Challenge Description}
\label{sec:description}

AMC 2025 consisted of three tracks, and in this section, we first introduce the task and the dataset in each track. Table~\ref{tab:datasets} summarizes the three tasks. Then, we describe the logistics of the challenge, including the timeline and rules. Table~\ref{tab:datasets} briefly summarizes the dataset used in each track.

\subsection{Track 1}

Track 1 focused on predicting human MOS for generated music in the MusicEval dataset \cite{musiceval}. The MusicEval dataset contains 2,748 mono audio clips, amounting to a total of 16.62 hours of audio, along with 13,740 ratings provided by 14 professional music experts, including 2 conservatory instructors and 12 advanced students. These clips were generated by 21 different TTM and TTA systems, spanning 31 models, in response to 384 diverse text prompts focused on classic and pop genres. The systems vary widely in terms of model size, year of development, accessibility, and commercialization status, capturing a broad spectrum of the generative music landscape.

Five experts rated each audio clip along two dimensions: overall musical quality and textual alignment, both on a 5-point Likert scale. The former captures the perceived musicality, rhythm, melody, and coherence, while the latter assesses the semantic relevance to the provided prompt. To support fair evaluation and model benchmarking, the dataset is randomly split into training, development, and test sets using a 70\%-15\%-15\% ratio. Robust quality control mechanisms were employed during the listening tests, including probe trials with human-created audio and consistency checks with duplicated clips, ensuring high reliability of the subjective ratings.

\subsection{Track 2}

Track 2 focused on predicting the four axes of Meta Audiobox Aesthetics~\cite{aes} for speech, music, and sound perceptual qualities, including Production Quality (PQ) on recording quality, Production Complexity (PC) on audio scenes, Content Enjoyment (CE) on listening experiences, and Content Usefulness (CU) for content creation. These dimensions were designed to provide detailed quality measurements and avoid ambiguity from a single MOS. Among the axes, PQ and PC are considered more objective, while CE and CU are considered more subjective. Meta worked with an external vendor to hire and train a pool of 158 annotators with audio-related backgrounds. Only annotators whose PQ and PC (the more objective axes) scores with a \textgreater~0.7 Pearson correlation coefficient with the expert-labeled ground truths in a curated golden set were qualified for AES-Natural annotations.

The training set for track 2 was the AES-Natural~\cite{aes} dataset, which included 950/1000/1000 speech/music/sound samples, respectively. Each sample was labeled by 10 well-trained annotators. Samples in AES-Natural had sampling rates varying from 16 to 48 kHz. Although all audio samples are sampled from public datasets, we asked participants to download these audio samples on their own because of the redistribution restrictions from the YouTube policies, etc.  

The test set for track 2 consisted of synthetic samples from TTS, TTA, and TTM systems.
For TTS, we used the test set of the LibriTTS-P dataset \cite{libritts-p}. To reflect the current trend in TTS research, we considered two types of TTS systems. The first type was prompt-based TTS, whose input consisted of an input text sequence and a natural language prompt that described the characteristics to be synthesized. The second type was reference-based TTS, whose input consisted of an input text sequence and a reference speech sample, such that the synthesized speech was expected to have the characteristics of the reference.
For TTA, we used the test set of the AudioCaps dataset \cite{audiocaps}, and for TTM, we used the MusicCaps dataset \cite{musiccaps}. Unless the model had a default setting, the duration of each TTA and TTM sample was ten seconds.
We first generated samples using the respective datasets and then filtered out samples shorter than four seconds. Finally, we randomly chose 85 samples per system, resulting in 3060 samples.

\subsection{Track 3}
\label{ssec:track3}

Track 3 focused on MOS prediction for synthesized speech at different sampling rates.  The dataset consists of four separate listening tests: one with only samples at a 16~kHz sampling rate, one at 24~kHz, one at 48~kHz, and one in which listeners rated all of the samples together.  Each of the four tests has a Part 1 and a Part 2, in each of which half of the listeners rated half of the samples.  The samples are English sentences including natural speech from LibriTTS-R \cite{koizumi2023libritts} and the Hi-Fi-CAPTAIN corpus \cite{hi-fi-captain}, as well as non-neural and neural vocoded speech \cite{morise2016world,yamashita2024fast}, TTS 
\cite{okamoto2025}, and super-resolutioned audio using AudioSR \cite{liu2024audiosr}.  All samples were normalized using pyloudnorm \cite{steinmetz2021pyloudnorm}.  There are 10 ratings per audio sample, and 20 listeners participated in total.

For the training phase, participants were provided with the Part 1 data from the 16~kHz, 24~kHz, and 48~kHz listening tests, which contained 120, 240, and 40 audio samples, respectively.  For the leaderboard, we used Part 1 of the mixed sampling rate listening test, so participants need to make predictions on the same set of 400 total audio samples but for a different listening test where all of these samples were evaluated together.  For the test phase, we used Part 2 of the mixed listening test, which also contained 400 samples, so the task is the same as the leaderboard task (to predict ratings when synthesized audio at different sampling rates appears together in the same test) but for a new set of previously-unseen utterances.

\subsection{Logistics}
\label{ssec:rules}

The challenge started with a training phase on April 5, and participants could use the provided training data to develop their systems. To promote the use of the leaderboard, except for track 2, we did not distribute the ground truth score labels of the development set, such that the participants must submit their scores to the publicly displayed leaderboard to check the model's performance on the development set.
The evaluation phase started on May 28, as we released the test set audio samples to the participants, giving them one week to prepare the final submission. Submissions were due on June 4, as we, at the same time, released the test set ground truth labels for participants to perform analysis and wrap up their paper. We also asked each team to submit a system description form based on the template we distributed. 

As usual, the challenge was held on CodaBench\footnote{\url{https://www.codabench.org/competitions/7029/}}, an open-source web-based platform for machine learning competitions and reproducible research. The participants could choose to participate in any track as long as they submitted their answers in the evaluation phase. Starting from VMC 2024, we established a new rule to restrict the data used for training to public datasets, with the exception that the participants are willing to make the resource open after the challenge (either making the previously in-house dataset publicly available or releasing the trained model checkpoints of their system). The same rule also applied in AMC 2025.

\begin{table}[t]

    \centering
    \caption{List of participant affiliations, in random order.}
    \label{tab:teams}

    \begin{tabular}{@{}l|ccc@{}}
        \toprule
        \multicolumn{1}{c|}{\multirow{2}{*}{\textbf{Affiliation}}} & \multicolumn{3}{c}{\textbf{Track}}   \\ \cmidrule(l){2-4} 
        & \textbf{1} & \textbf{2} & \textbf{3} \\ \midrule
        National Taiwan Normal University                 & V          & V          &            \\
        Unknown (1)$^\dagger$                                               & V          &            &            \\
        Fusic Co.,Ltd., Japan                                     & V          &            & V          \\
        \begin{tabular}[l]{@{}c@{}}Nanyang Technological University, Singapore\\ \& National Taiwan University  \end{tabular}               & V          &            &            \\
        Xiaomi, China                                             & V          &            &            \\
        Michigan State University, USA$^\ddagger$                            & V          &            &            \\
        Alibaba Inc., China                                       & V          &            &            \\
        Hokkaido Denshikiki Co., Ltd., Japan                      & V          &            &            \\
        Ningbo University, China                                  & V          &            &            \\
        National Taiwan University (1)                    &            & V          &            \\
        ByteDance, China (1)                                      &            & V          &            \\
        ByteDance, China (2)                                      &            & V          &            \\
        Unknown (2)$^\dagger$                                               &            & V          &            \\
        The University of Tokyo, Japan (1)                        &            & V          &            \\
        CyberAgent Inc., Japan                                    &            & V          &            \\
        Academia Sinica                                   &            & V          &            \\
        THAU group, Spain                                         &            & V          &            \\
        Tianjin University, China$^\ddagger$                                 &            & V          &            \\
        The University of Tokyo, Japan (2)                        &            &            & V          \\
        Zhejiang University, China                                &            &            & V          \\
        Kyoto University, Japan                                   &            &            & V          \\
        Samsung Electronics, Greece                               &            &            & V          \\
        AKCIT Federal University of Goias, Brazil                 &            &            & V          \\
        National Taiwan University (2)                    &            &            & V          \\
        \bottomrule
        \multicolumn{4}{l}{
            $^\dagger$: Did not submit their system description form.
        } \\
        \multicolumn{4}{l}{
            \begin{tabular}[l]{@{}l@{}}
                $^\ddagger$: Did not submit their system description form, but we\\inferred their affiliation from their email address.
            \end{tabular}
        } 
    \end{tabular}
\end{table}

\section{Participants and baseline systems}
\label{sec:participants}

Table~\ref{tab:teams} shows the participants, their affiliations, and the tracks in which they participated.
In total, we received evaluation phase submissions from 24 unique teams, which was the most among the previous VMCs, demonstrating the increasing interest in audio quality assessment. 
While we identified some returning participants from the previous VMCs, most teams were new to the challenge series, indicating that our new direction indeed attracted newcomers.
Participants spanned from countries and regions worldwide, with 14 teams from academia, eight from industry, and two unknown.
For each track, we had nine, ten, and seven participating teams, respectively, with only two teams participating in multiple tracks. From the feedback in the system description forms, most participants expressed their interest in all three tracks, but chose to focus on one track due to time and computational constraints. As in previous VMCs, we individually informed each team of their randomly-assigned team ID, and in the rest of this paper, we will refer to each team by their ID.

The baseline system for track 1\footnote{\url{https://github.com/NKU-HLT/MusicEval-baseline}}, noted as B01, was based on a pretrained CLAP model \cite{laionclap2023, htsatke2022}, which utilized an audio encoder from HTSAT and a text encoder from RoBERTa. Two separated 3-layer multi-layer perceptron (MLP) regression heads were applied to predict overall musical quality and textual alignment scores after extracting text and audio embeddings from the generated music sample and its corresponding text description. The model was trained with an L1 loss across the two dimensions, with a learning rate of 0.0005 and a batch size of 64. 

The baseline system for track 2\footnote{\url{https://github.com/facebookresearch/audiobox-aesthetics}}, noted as B02, was based on a pretrained WavLM~\cite{wavlm} followed by MLP blocks to predict the four axes. The WavLM model consisted of 12 Transformer~\cite{transformer} layers with 768 hidden dimensions, and a learnable weighted sum layer was used to aggregate feature maps from all layers to form the audio embeddings. Then, four MLP blocks independently processed the audio embeddings to predict the four scores, and each block consisted of five linear layers, layer normalization~\cite{layer_norm}, and GeLU activation~\cite{gelu}. The training objective included a mean absolute error (MAE) loss and a mean squared error (MSE) loss. All the training data was resampled to 16 kHz mono audio, in order to be compatible with the pretrained WavLM. Notably, the model was trained on a 500-hour in-house dataset from Meta.

The baseline system for track 3\footnote{\url{https://github.com/nii-yamagishilab/mos-finetune-ssl}}, noted as B03, was based on a pretrained SSL-MOS model \cite{cooper2022generalization}, fine-tuned on the data from the three listening tests in the training dataset combined, for 50 epochs.  Since SSL-MOS only accepts 16~kHz input, we downsampled all audio before running the prediction.  While this process removes salient information at higher frequencies that is likely to affect listener judgments, this baseline tells us the extent to which listener preferences can be predicted from the lower-frequency information only.

\section{Results}
\label{sec:results}

\subsection{Evaluation metrics}

Following previous VMCs, for each evaluation dimension, we computed both utterance-level and system-level mean squared error (MSE), linear correlation coefficient (LCC), Spearman rank correlation coefficient (SRCC), and Kendall's Tau rank correlation coefficient (KTAU).  The primary metric was also aligned with previous challenges, system-level SRCC. The reason was that when it comes to evaluating audio generation systems, we mainly want to know the \textit{rankings} of the systems under consideration. Note that for track 3, instead of ``systems,'' we considered ``conditions'' as unique combinations of the speech generation method and sampling rate. Due to space limits, in this paper, we will only show figures related to the main metrics, and the raw scores and the rankings of each team will be shown on the challenge website. 

\begin{figure}[t]
	\centering
	\includegraphics[width=\linewidth]{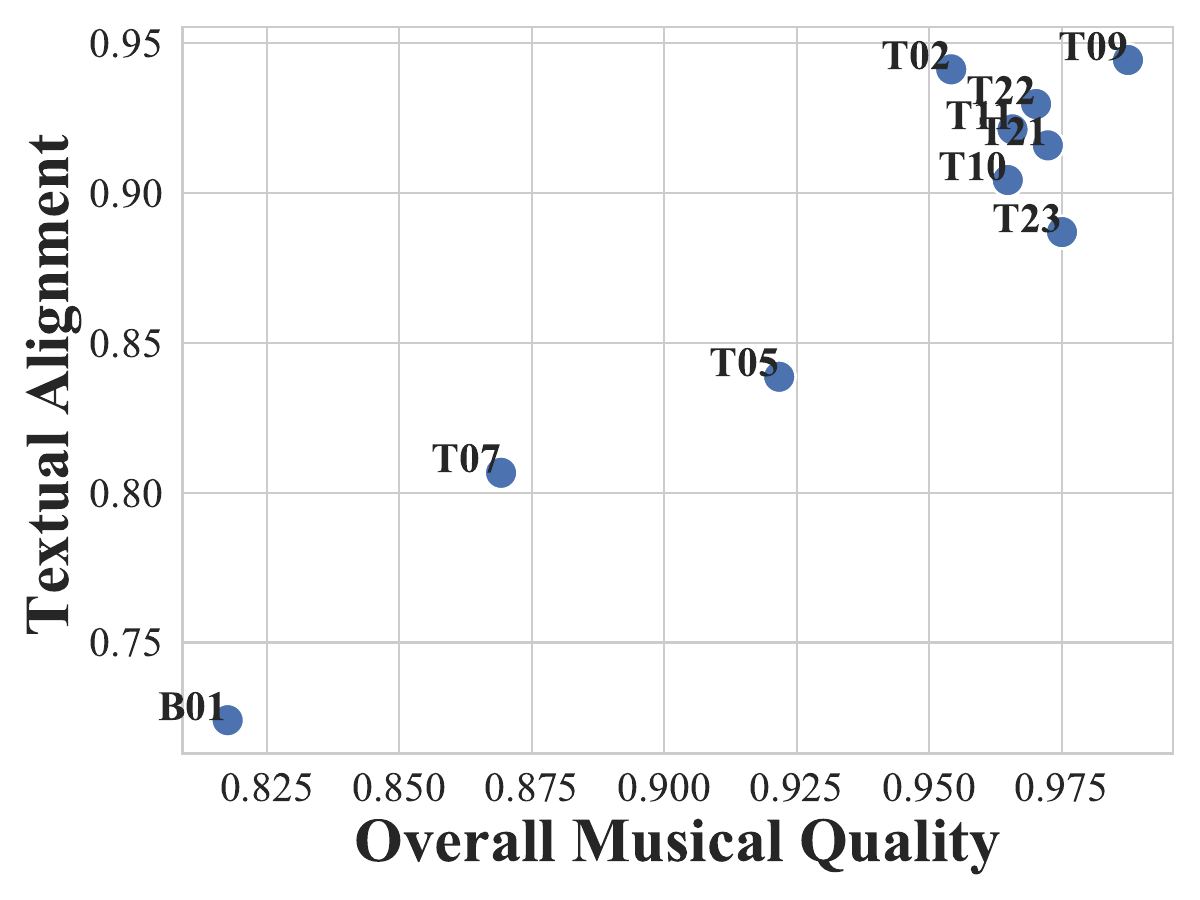}
	\caption{\label{fig:track1-scatter}Scatter plot of system-level SRCC values of all participants in track 1.}	
\end{figure}
\begin{figure}[t]
	\centering
	\includegraphics[width=\linewidth]{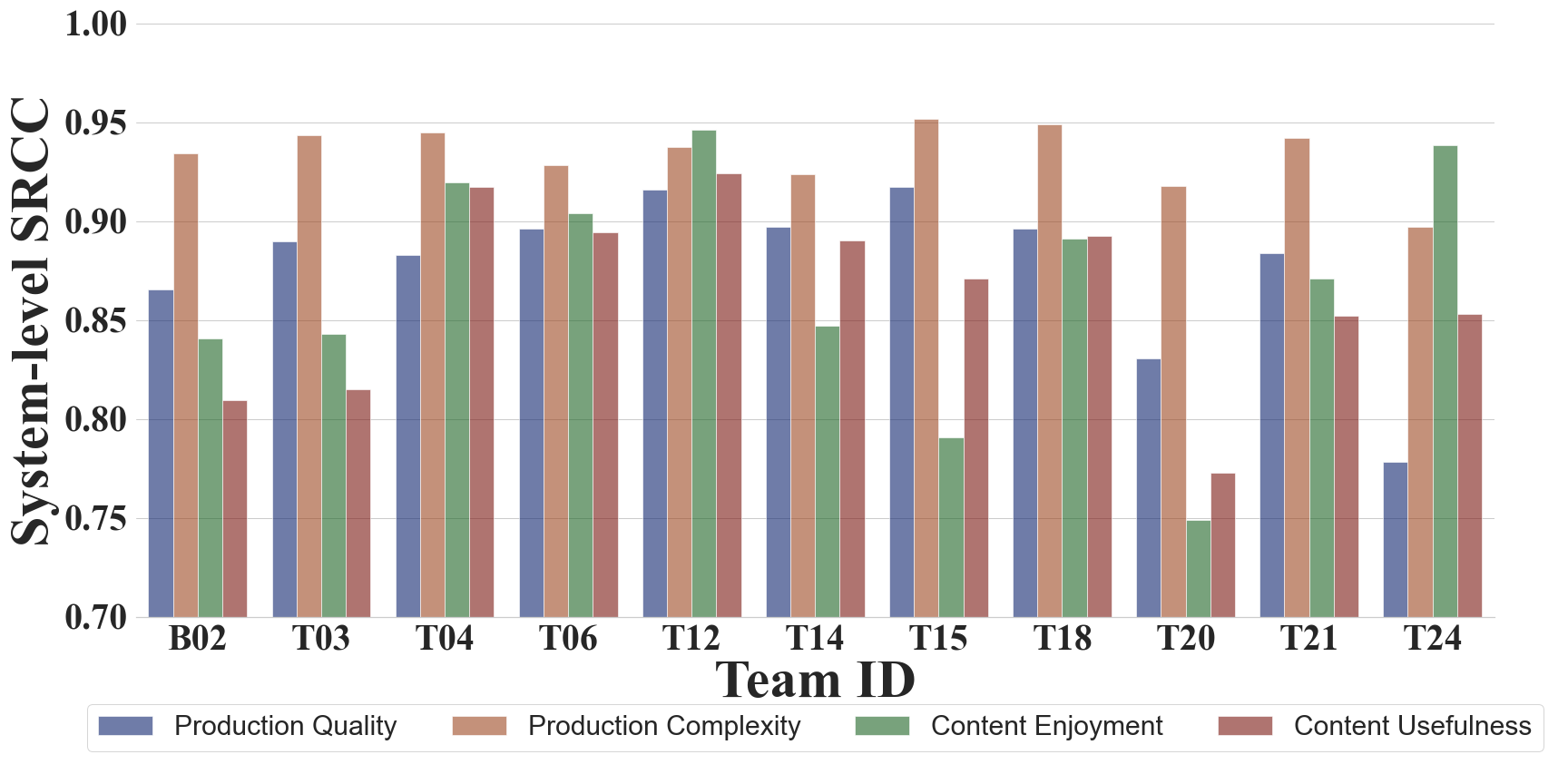}
	\caption{\label{fig:track2-bar}Bar plot of system-level SRCC values of all participants in track 2.}	
\end{figure}
\begin{figure}[t]
	\centering
	\includegraphics[width=\linewidth]{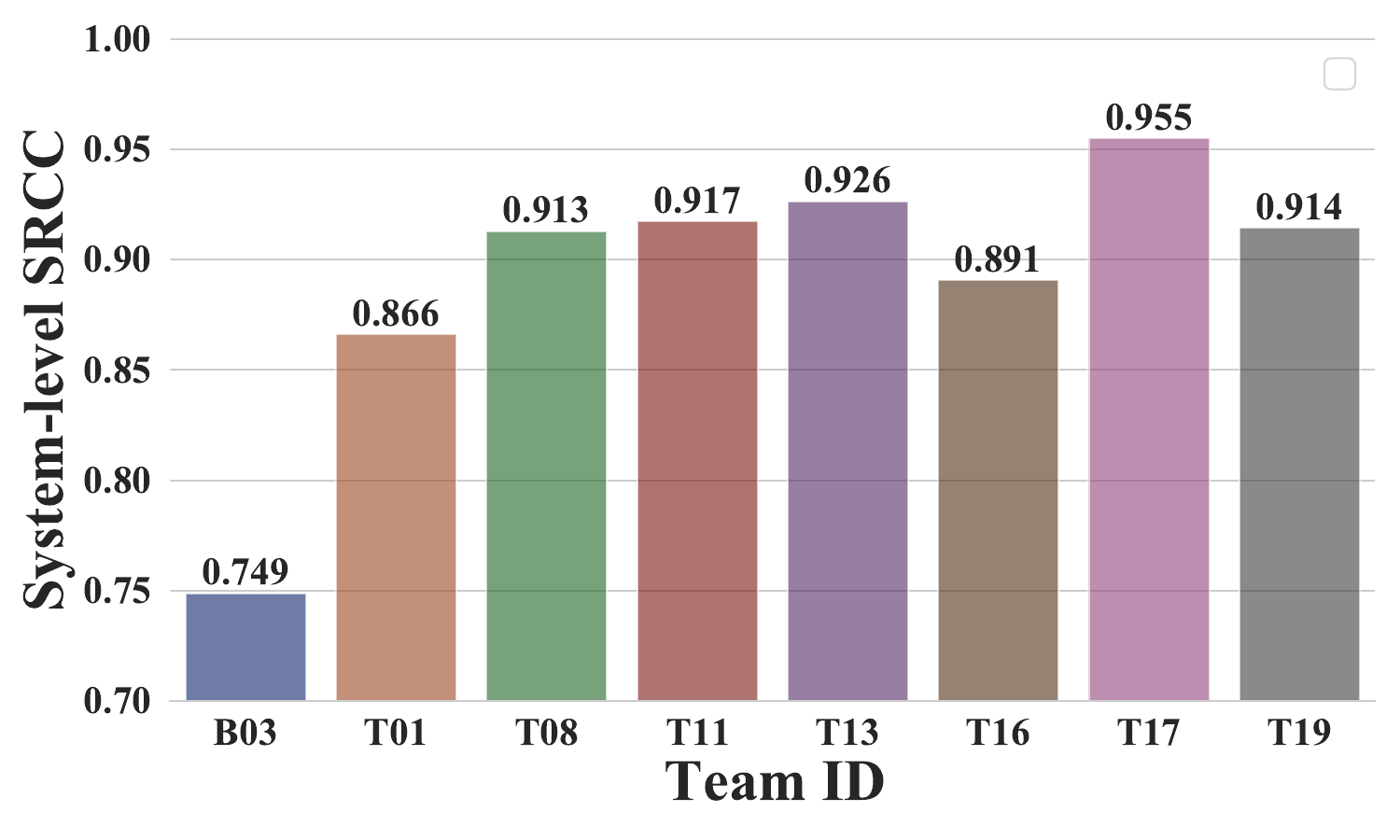}
	\caption{\label{fig:track3-bar}Bar plot of system-level SRCC values of all participants in track 3.}	
\end{figure}

\subsection{Track 1 results}
Figure~\ref{fig:track1-scatter} shows the system-level SRCC scores for all participants in Track 1. Among them, system T09 stood out as the best overall in both musical quality and textual alignment. Specifically, it surpassed the baseline system by 20.8\% and 30.4\% in the two dimensions, respectively. Following closely, a group of strong performers, including T02, T22, T11, T21, and T10, exhibited well-balanced results. T23 also demonstrated competitive performance. In contrast, systems like T05 and T07 showed potential but lagged behind the leading group. Notably, the baseline system B01 ranked the lowest, highlighting the substantial improvements made by participating teams.


Beyond system-level improvements, it is equally important to examine broader trends and trade-offs across the two evaluation dimensions.
Systems that performed well in one dimension tended to excel as well in the other dimension, but the relationship was not strictly linear. Within the top-performing group, such as T02, T09, T10, and T22, some systems achieved higher scores on musical quality, while others led in textual alignment. This suggests that shared feature encoders contributed to consistent performance, while task-specific output heads allowed partial decoupling. In terms of difficulty, textual alignment scores were consistently lower than musical quality across most systems. For example, T10 scored 0.965 in overall music quality and 0.902 in textual alignment, indicating that textual alignment was generally harder to model due to its reliance on cross-modal understanding.

\subsection{Track 2 results}

Figure~\ref{fig:track2-bar} shows a bar plot of the system-level SRCCs for all systems in track 2 w.r.t. the four axes in Audiobox Aesthetics.
The baseline was outperformed by most of the teams: the system-level SRCC of B02 ranked 9/7/9/10 in PQ/PC/CE/CU, respectively. We also observed that in terms of the system-level SRCC, no team ranked first in all axes.
Notably, from the system description analysis which will be presented in Section~\ref{sec:system-description-forms}, compared to the baseline system which was trained on 500 hours of data with the Audiobox Aesthetics labels, most teams trained their system using only the AES-natural dataset, which somewhat indicates that scaling up the training data is not essentially effective.


\subsection{Track 3 results}

Figure~\ref{fig:track3-bar} shows a bar plot of the system-level SRCCs for the teams that participated in the task of predicting MOS for synthesized speech at different sampling rates. 

Following methods from the VoiceMOS Challenge 2024, we used absolute differences between predicted and ground-truth rank of each synthesis condition (where a ``condition'' is a combination of synthesis system and sampling rate) in Track 3 to determine whether some conditions tended to be more difficult to predict than others.  Additionally, we analyzed whether any sampling rate was more difficult to predict.

Considering sampling rates, we first accumulated ranking errors made by all predictors for all synthesis conditions of each sampling rate, and then normalized by number of synthesis systems that used that sampling rate, since the number of systems for each sampling rate was different (4 systems for 16 kHz, and 8 systems for both 24 kHz and 48 kHz).   Ranking errors were computed by taking the absolute value of the difference between the ground-truth ranking of a condition and a team's predicted ranking.  In addition to computing aggregated ranking errors, we also computed absolute score errors by taking the absolute value of the difference between a condition's ground-truth MOS value and the predicted one.  By both the absolute ranking and score difference metrics, the 16 kHz systems had the most errors.  Although the development of MOS prediction for 16 kHz audio by itself has strong precedent in past VMCs and other works, we can observe that the task becomes more difficult when audio at other sampling rates is included.

In order to determine which conditions were the most difficult to predict, for each predictor, we computed the absolute ranking error of each condition.  Then, we counted how many teams had each condition in their top 5 most difficult-to-predict ones.  In this way, we identified the most difficult conditions as: (1) natural speech down-sampled and then super-resolutioned to 24 kHz, which all eight predictors had as one of their top-5 most difficult conditions, (2) vocoded speech from a neural vocoder at 16 kHz, which was difficult for 7 of the predictors, and (3) natural speech at 16 kHz, which was also difficult for 7 predictors.  Two of the most difficult conditions are 16 kHz ones, which track with our earlier finding that the 16 kHz sampling rate was more difficult overall.  Looking into the direction of the errors, we observed that for the two most difficult 16 kHz conditions, all eight predictors consistently predicted worse ranks for those than their ground-truth ranks, and seven teams also predicted better ranks for the super-resolutioned 24 kHz natural speech than its actual ranking.

\section{Insights from the system description forms}
\label{sec:system-description-forms}

We received system description forms from all teams except for T05, T10, T14, and T20. Below, we summarize insights gained from the submitted forms.

\subsection{Overall analysis of the participating systems}
\label{ssec:system-analysis}

\subsubsection{Data}

First, we observed that only three teams used additional datasets, while the others used solely the training and development sets we provided. This is possibly due to the nature of the tracks: there exist almost no other datasets that follow the evaluation axes of tracks 1 (musical overall quality and textual alignment) and 2 (the four axes in Meta Audiobox Aesthetics), thus making it difficult to make use of additional data.

Among the three teams, two teams used the labels of the PAM dataset \cite{pam}, which were publicly available on the Meta Audiobox Aesthetics github repo. In addition, the BVCC \cite{bvcc} and EARS \cite{ears} datasets were also used, mainly for model pre-training. Notably, T12, the top-performing team in track 2, used the additional PAM and BVCC datasets to perform semi-supervised training. See details in Section~\ref{sssec:t12}.

\subsubsection{Self-supervised learning models}

Self-supervised learning (SSL) has become an indispensable ingredient in modern-day speech processing tasks, and this year's challenge was no exception. The most popular SSL model choices were CLAP, WavLM, and wav2vec 2.0, since they were already used in the baselines. Another popular category was general audio encoders, including Qwen-Audio \cite{qwen-audio}, BEATs \cite{beats}, M2D \cite{m2d}, EnCodec \cite{encodec}, and Dasheng \cite{dasheng}. In addition, to tackle the unique nature of track 1, participants explored music-specific SSL models like MERT \cite{mert} and MuQ \cite{muq}, text SSL models like BERT \cite{bert}, RoBERTa \cite{roberta} and Qwen-3.0 \cite{qwen3}. Finally, some participants also employed other speech SSL models, including Whisper \cite{whisper}, HuBERT \cite{hubert}, MMS \cite{mms} and EAT \cite{eat}.

\subsubsection{Training techniques} We categorize and enumerate techniques employed by participants below.

\noindent\textbf{Input features}: While most teams explored the use of one or multiple SSL representations, T17, the top-performing team in track 3, included Mel-spectrograms and Mel-frequency cepstrum coefficients (MFCCs) as input. To balance between multiple input sources, several teams tried to learn a set of dynamic weights for each input. Notably, in track 3, four out of the seven teams used a discrete sampling rate ID as input. Finally, although listener modeling \cite{ldnet} proved to be effective in previous VMCs, only two teams used such a technique this year.

\noindent\textbf{Models}: Many teams tried modern day model architectures, including KANs \cite{kan}, Mamba2 \cite{mamba2}, and sampling frequency independent convolution layers \cite{sampling-rate-independent}.

\noindent\textbf{Training objectives}: In this year's challenge, we observed a diverse range of loss functions, including triplet loss, Huber loss, Sinkhorn optimal transport loss, cross-entropy loss with smoothing, rank-consistent ordinal regression loss, and pairwise adaptive margin ranking loss.

\noindent\textbf{Model ensemble}: We observed a strong relationship between the eight out of 24 teams that used model ensembling and their final performance, which was also present in previous VMCs. Notably, the top-performing systems in each track all used this technique. 

\subsection{Brief descriptions of the top performing systems}
\label{ssec:top-systems}

\subsubsection{\textbf{T09}, top performing system in track 1}

T09 ranked first on 13 out of the 16 metrics in track 1. Their base model is based on the MuQ and the RoBERTa features. For the overall musical quality axis, a Transformer block takes the MuQ features as input, followed by an attention pooling layer and a fully connected layer to predict the score. For the textual alignment axis, a cross-attention Transformer block takes both MuQ and RoBERTa features as input, again followed by an attention pooling layer and a fully connected layer to predict the score.

T09 also performed stacking-based ensemble learning. First, they prepared nine models. First, five base models with different random seeds were trained using a smoothed cross-entropy loss. Then, two base models were trained using a rank-consistent ordinal regression loss. Finally, two slightly modified models were trained. After the nine models were trained, a ridge regression model was trained to aggregate the predictions of the nine models.

\subsubsection{\textbf{T12}, top performing system in track 2}
\label{sssec:t12}

T12 ranked first on 17 out of the 32 metrics in track 2. On system-level SRCC, they ranked first on CE and CU, and ranked second on PQ. T12 performed model ensembling with two models. The first model is an improved version of the baseline, where they replaced the MLP layers with GR-KAN \cite{gr-kan}, a variant of KANs. In the second model, they utilized VERSA \cite{versa}, a recent proposed suite of audio evaluation metrics, and calculated scores from 28 non-intrusive metrics. These scores are then sent into an XGBoost-based regressor \cite{xgboost}. Both models predict values of PQ, PC, CE, and CU. The final scores are calculated by an ensemble model with four KAN-based models and a VERSA-based model. They also performed semi-supervised learning using the noisy student training technique \cite{noisy-student} on the PAM and the BVCC datasets.

\subsubsection{\textbf{T17}, top performing system in track 3}

T17 ranked first on five out of the eight metrics in track 3. They only used the provided training and development sets. Their base model is an extension of the SSL-MOS baseline with three additional inputs, resulting in four input features: the SSL features, sampling rate ID, Mel-spectrogram, and MFCC. To reflect different input sampling frequencies, they employed a multi-scale convolutional block for the mel-spectrogram. The output of the four streams are then passed through a bidirectional long short term memory layer, followed by a fully connected layer to predict the score. The training objective included an MAE loss, a ranked-based loss, and a correlation loss. They ensembled three models: the base model trained with the complete training set, the base model trained with part of the training set, and a model without the MFCC feature trained on the complete training set.

\subsection{Feedback from the participants}
\label{ssec:feedback}

Most feedback from participants was positive, complementing the design of the tracks, convenience of the platform and the fast response from the organizers. Negative comments include ambiguous instructions and rules, a short training phase, scarce data, overly strong baselines, and page limits for the challenge papers. Among them, the most common comment was the inconvenience when retrieving the track 2 training set. Since a part of the training samples of track 2 were based on YouTube videos, as organizers we could not redistribute them due to YouTube policies. We will try to avoid using such data sources in the future.

Tasks that the participants wished to see in the future include (1) more speech types, including multi-lingual speech, expressive TTS, prompt-based TTS, and non-verbal speech; (2) multi-dimensional music evaluation, including music rhythm, music theory, music diversity, and music style similarity; (3) other audio types, including ambient sound and video-to-audio generation; (4) other topics, including preference score estimation and real-time evaluation.

\section{Conclusion}

This paper summarized the AudioMOS Challenge 2025. Among the 24 participating teams, we observed that the use of advanced SSL representations and model architectures, paired with well-designed training objectives and ensemble learning techniques, was effective in the tracks this year. 
We believe the datasets and the insights brought by the challenge can further benefit the progress in the field of audio generation evaluation.

\section*{Acknowledgment}
We thank Dr. Takuma Okamoto from the National Institute of Information and Communications Technology, Japan, for curating and sharing the dataset used in track 3. This work was partly supported by JST AIP Acceleration Research JPMJCR25U5, Japan.

\bibliographystyle{IEEEtran}
\bibliography{ref}

\end{document}